\documentclass[12pt]{article}
\usepackage{graphics,epsfig}
\usepackage[english]{babel}
\begin{document}

\begin{titlepage}

\begin{center}

\vglue2.5cm

\raisebox{0.5cm}[0cm][0cm] {
\begin{tabular*}{\hsize}{@{\hspace*{5mm}}ll@{\extracolsep{\fill}}r@{}}
\begin{minipage}[t]{3cm}
\vglue.5cm
\end{minipage}
&
\begin{minipage}[t]{5cm}
\vglue.5cm
\end{minipage}
&
\begin{minipage}[t]{5cm}


\end{minipage}
\end{tabular*}
}
\end{center}

\vskip3.5cm
\begin{center}

{\Large{\bf The LUCIAE model predictions \\
on the light vector meson production in 
\\ proton-proton,
proton-nucleus and nucleus-nucleus  \vspace{0.2cm} \\ collisions
at 158 A GeV}}

\vspace{3.5 cm}

{\large  M. Atayan, H. Gulkanyan}

\vglue-5mm

\vspace{1.0 cm}

\noindent { Yerevan Physics Institute, Armenia}
\end{center}

\vglue 2cm

\begin{abstract}

\noindent The LUCIAE model predictions are compared with
experimental data on light vector meson ($\rho,\omega,\phi$)
production in pp, p+Pb and Pb+Pb collisions at 158 A GeV.
The model reproduces the general trends of the data,
but fails in describing the yield of high-$p_T$
vector mesons in the central Pb+Pb collisions. Predictions for
In+In and p+A (A varying from Be to U) collisions at 158 A GeV
are also presented.

\end{abstract}

\end{titlepage}

\newpage

\section{Introduction}

The enhancement of the strange quark production in heavy ion
collisions, in particular, the enhanced yield of $\phi$ meson
relative to non-strange $\rho$ and $\omega$ mesons is suggested as
a possible signature of quark-gluon plasma (QGP) formation
\cite{ref1,ref2}. On the other hand, such an enhancement is also
predicted by conventional (no QGP) models, in particular, by the
LUCIAE \cite{ref3}, an updated version of the FRITIOF model
\cite{ref4,ref5}. \\ The first and most detailed experimental data
on the light vector meson ($\rho,\omega,\phi$) production in
nuclear collisions are obtained at the CERN SPS. However, up to
now no attempt was undertaken to describe simultaneously these
data in the framework of a unique approach. \\ In this Note, the
LUCIAE model is employed (with somewhat updated parameters) to
compare the theoretical predictions with the existing data on
$\phi$ and ($\rho + \omega$) production in pp, p+Pb and Pb+Pb
collisions at 158 A GeV \cite{ref6,ref7}. In Section 2, the
parameters governing the strange particle yield and the transverse
momentum distribution of produced hadrons are described. Section 3
is devoted to the comparison of the model predictions with
experimental data. Predictions for In+In and p+A 
(A varying from Be to U) collisions are also presented. 
The results are summarized in Section 4.

\section{The model parameters}

The choice of the LUCIAE model parameters governing the
multiplicity and transverse momentum distributions of final
particles produced in nucleon-nucleon (hadron-nucleon) collisions
at SPS energies is motivated in \cite{ref8} from the comparison
with experimental data. The following values of parameters, some
of which differ from the default ones \cite{ref5}, will be used
below:
\\ - The minimum transverse momentum $q_{Tmin}$ transferred during
the (semi)hard scattering of partons of the colliding hadrons is
$q_{min}\equiv$ VFR(12) = 0.6 GeV$/c$;
\\ - The width of the Gaussian distribution of the soft transverse
momentum transfer $Q_T$ between the two colliding hadrons
(strings) is given by $<Q_T^2> \equiv$ VFR(6) = 0.1 (GeV$/c)^2$;
\\ - The width of the Gaussian distribution of the primordial transverse
momentum $Q_{2T}$ carried by the string ends (valence quark or
diquark) is given by $<Q^2_{2T}> \equiv$ VFR(7) = 0.1 (GeV$/c)^2$;
\\ - The width of the Gaussian distribution of the transverse
momentum ($p_x$ and $p_y$ in the string reference frame) acquired
by hadrons as a results of the string fragmentation is ${\sigma}_x
= {\sigma}_y\equiv$ PARJ(21) = 0.405 GeV$/c$, equal to the default
value \cite{ref5}. \\ In the LUCIAE model, the broadening of the
transverse momentum $p_T$ distribution in nuclear collisions is
suggested to be caused by a novel ('firecracker') mechanism for
gluon emission by a high-density collective state formed by
several overlapping strings \cite{ref9}. Besides, a
phenomenological mechanism is introduced for the enhancement of
the strange quark yield in the fragmentation of the collective
string states \cite{ref10}.

\noindent The model basic parameters governing the yield of
hadrons with strange content are: PARJ(2) = P(s)/P(u), the rate of
the ($s \bar{s}$) production compared to ($u \bar{u}$) or ($d
\bar{d}$) production, and PARJ(3) = (P(us)/P(ud))/(P(s)/P(d)), the
extra suppression of the strange diquark production compared to
the normal suppression of strange quarks. \\ For 200 GeV pp
collisions, these parameters are tuned in \cite{ref10} by
comparison with data on the strange particle production: PARJ(2) =
0.2, PARJ(3) = 0.3. At these values, however, the model prediction
\cite{ref11} for the $\phi$ meson average multiplicity
$<n_{\phi}>$ in 158 GeV pp collisions  somewhat (by 18\%)
overestimates the experimental value \cite{ref6}. Below (in
Section 3) we will, therefore, use for 158 GeV pp collisions a
slightly smaller value for PARJ(2) = 0.19, keeping the value of
PARJ(3) = 0.3.

\noindent Below two options of the model are considered: without
and with the final state interactions(FSI).The latter includes the
following inelastic reactions involving $\rho$ and/or $\omega$
mesons: $ {\pi} N {\rightleftharpoons} {\rho} N$, ${\bar{N}N}
\rightarrow {\rho\omega}$ and ${\bar{Y}N}\rightarrow
K^{\ast}{\omega}$ $(Y = \Lambda~or~\Sigma)$, leading to an
increased yield of $\omega$, especially in heavy ion collisions
(however, practically not influencing the yield of $\rho$), while
the FSI effects on the $\Phi$ meson production are neglected
\cite{ref3,ref11}. To trace the predicted FSI effects concerning
the yield of $\rho$ and $\omega$ , the
predictions are presented below for both model options.\\
\noindent Note finally, that in the model predictions the
following branching ratios for $\phi, ~\rho$ and $\omega$ dimuon
decay are used: $B_{\phi}^{\mu\mu} = (2.87\pm 0.22) \cdot
10^{-4}$, $B_{\rho}^{\mu\mu} = (4.60\pm 0.28) \cdot 10^{-5}$
\cite{ref12} and $B_{\omega}^{\mu\mu} \approx B_{\omega}^{ee} =
(7.07\pm 0.19) \cdot 10^{-5}$ \cite{ref13}. The quoted errors of
the branching ratios are taken into account in calculations.

\section{The model predictions and comparison with experimental data}

In Table 1 the model predictions for the total average
multiplicity~ $<n_{\phi}>$ ~and ~the $<n_{\phi}>/<n_{\pi}>$ ratio
are compared with the NA49 data \cite{ref6}. An agreement is
seen both for pp, p+Pb and central (at $b<3.5~fm$) Pb+Pb data.
The predictions for 158 A GeV In+In collisions are also presented.

\begin{table}[ht]
\begin{center}
\begin{tabular}{|c|c|c|}
  \hline



Reaction& $<n_{\phi}>$& $<n_{\phi}>/<n_{\pi}>$\\ \hline pp& &
\\ NA49& $(1.20\pm0.15) \cdot10^{-2}$ &$(0.42\pm0.05) \cdot10^{-2}$ \\
LUCIAE&$(1.39\pm0.04) \cdot10^{-2}$ &$(0.46\pm0.02) \cdot10^{-2}$
\\ \hline
p+Pb& &
\\ NA49& $-$  &$(0.70\pm0.09) \cdot10^{-2}$ \\
LUCIAE&$(4.7\pm0.2) \cdot10^{-2}$ &$(0.75\pm0.03) \cdot10^{-2}$
\\ \hline
In+In& &\\ LUCIAE&$0.85\pm0.01$ &$(0.94\pm0.01) \cdot10^{-2}$
\\ \hline
Pb+Pb& &\\ (central)& &\\ NA49& $7.6\pm1.1$ &$(1.24\pm0.18)
\cdot10^{-2}$ \\ LUCIAE&$7.00\pm0.04$ &$(1.13\pm0.01
)
\cdot10^{-2}$
\\ \hline
\end{tabular}
\end{center}
\caption{The multiplicity  $<n_{\phi}>$ and the ratio
$<n_{\phi}>/<n_{\pi}> = 2 <n_{\phi}>/<n_{{\pi}^+} +n_{{\pi}^-}>$}
\end{table}

\noindent The model approximately reproduces the pp data \cite{ref6} on 
the $\phi$ meson rapidity (Fig.~1) and transverse mass $M_T$ (Fig.~2)
distributions. The agreement with data for the both distributions
is somewhat better than in the original paper \cite{ref11},
where the predictions for the $\phi$-production are presented for
the first time in the framework of the LUCIAE model, with somewhat
different set of parameters mentioned above in Section 2.\\
 \noindent The comparison with data for Pb+Pb central
collisions \cite{ref6} is also shown in Figs.~1 and 2. The model
overestimates by $\sim$20\% the data at mid-rapidity and predicts
somewhat narrower y-distribution (with width ${\sigma}_y = 0.97\pm
0.02$), than the experimental one (${\sigma}_y^{exp} = 1.22\pm
0.16$). The inverse slope of the predicted $M_T$- distribution,
parameterized as $dn_{\phi}/dM_T^2dy \sim exp(-M_T/T_{\phi})$,
$T_{\phi} = 232\pm$ 2 MeV, is significantly smaller than the
measured one $T_{\phi}^{exp} = 305\pm$ 15 MeV. The recent attempt
\cite{ref14} to reproduce the $M_T$- distribution is also
unsuccessful: the LUCIAE predictions prevail the data at small
$M_T-m_{\phi} <$ 0.15 GeV/$c^2$, while at large $M_T-m_{\phi}
>$ 0.5 GeV/$c^2$ the predicted values are lower than the experimental ones.
\\ This inconsistency at large $M_T
>$ 1.5 GeV$/c^2$ is more significant when comparing the model
predictions with the NA50 data for the dimuon channel of $\phi$
production in Pb+Pb collisions \cite{ref7}. As seen from Fig.~3
(the left panel), the model predictions are 1.5-2 times lower of
the data on the differential cross section integrated over the
almost whole domain of impact parameter (except for the most
peripheral collisions). The predicted value of $T_{\phi} = 220\pm$
7 MeV is, however, consistent with the measured one
$T_{\phi}^{exp} = 228\pm$ 10 MeV. Better agreement is obtained for
the $\rho + \omega$ differential cross section (Fig.~3, the right
panel). Note, that the predicted values of the cross-sections (and
of the multiplicities $<n_{\phi}>_{\mu\mu}$ and $<n_{{\rho} +
{\omega}}>_{\mu\mu}$, see below) are multiplied by a factor
$A_{CS}$ = 0.5, equal to the NA50 setup acceptance for the
Collins-Soper angle. In this acceptance, the predicted inclusive
cross section $\sigma_{\rho+\omega}^{\mu\mu}$ (at $1.5<M_T<3.5$
GeV$/c^2$ and $0<y_{cm}<1$) is equal to $191\pm6{\mu}b$ for the
model option without the FSI, while, owing to the introduction of
the latter, this value reaches $240\pm8{\mu}b$, consistent with
the measured one.\\ The dependence of the vector meson yield on
the number $N_{part}$ of participant nucleons in Pb+Pb collisions
and that per participant nucleon, $<n_{\phi}>_{\mu\mu}/N_{part}$
and $<n_{{\rho}+{\omega}}>_{\mu\mu}/N_{part}$, are plotted in
Fig.~4. It is seen, that the model predicts an increasing yield
per participant nucleon with increasing $N_{part}$, owing to the
collective string mechanism incorporated into the model. However,
this enhancement for $\phi$ is much moderate than observed in the
experiment \cite{ref7}: the model predictions are lower of the
data by a factor about 3 for the most central collisions (Fig.~4,
the top panels). It should be stressed, that the model fails not
only for the hidden strangeness sector, but also for the dimuon
channel of the high-$M_T$ $\rho+\omega$ production. As seen from
Fig.~4 (the bottom panels), the model predicts about 1.8 times
smaller ${<n_{{\rho} + {\omega}}>}_{\mu \mu}$, than measured in
the most central collisions, while due to the introduction of the
FSI this discrepancy reduces to about 1.4.

The model predictions concerning the $\phi$, $\rho$ and $\omega$
production in 158 A GeV In+In collisions for the acceptance of the
NA50/60 setup ($0<y_{cm}<1$ and $A_{CS}$ = 0.5) are presented in Figs.~5
and 6. In this acceptance, the integrated over $M_T$ cross sections
are predicted to be  $\sigma_{\phi}^{\mu\mu}$=0.21mb,
 $\sigma_{\rho}^{\mu\mu}$=0.57mb and $\sigma_{\omega}^{\mu\mu}$=
0.83mb for the option without FSI, while
$\sigma_{\rho}^{\mu\mu}$=0.55mb and $\sigma_{\omega}^{\mu\mu}$=
1.09mb when the FSI is taken into account. It is seen from Fig.~5,
that the $N_{part}$ - dependence of $<n_{\phi}>$ and
$<n_{\omega}>$ (when the FSI is taken into account for the latter)
is significantly stronger, than that for $<n_{\omega}>$ (when the
FSI is not taken into account) and for
$<n_{\rho}>$. The inverse slope parameter T (plotted in Fig.~6)
for $\phi$ is practically the same as for $\rho$ and $\omega$,
when the FSI is taken into account for the latters.

The predictions for p+A collisions (A $\equiv$ Be, Al, Cu, In, W,
Pb, U) are presented, at $0<y_{cm}<1$ and $A_{CS}$ = 0.5, in
Figs.~7 and 8. Again, owing to the collective string mechanism,
the predicted mean multiplicity  $<n_{\phi}>$ increases with A
faster, than  $<n_{\rho}>$ and $<n_{\omega}>$. The ratio
$<n_{\phi}>/<n_{\omega}>$ for the heaviest target (U) exceeds by
about 1.5 times that for pp collisions. The parameter A tends to
increase with A (Fig.~8), being for $\phi$ systematically higher
as compared to that for $\rho$ and $\omega$.

\section{Summary}
Predictions of the LUCIAE model, an updated version of the FRITIOF event
generator, for the $\rho, \omega, \phi$ production
in proton-proton, proton-nucleus and nucleus-nucleus collisions at
158 A GeV are presented and compared with the available experimental data.
The model parameters used for pp collisions were preliminary tuned by
comparison with the data on the multiplicity distribution and
inclusive spectra of charged particles and the strange particle
yield in proton-proton (hadron-proton) interactions at the SPS
energies.
\\ The model predictions agree with the data \cite{ref6} on the
total average multiplicity $<n_{\phi}>$ and the ratio of
$<n_{\phi}>/<n_{\pi}>$ for pp, p+Pb and central Pb+Pb
collisions, as well as approximately reproduces the
$M_T$ and y - distributions of  $\phi$ meson in pp collisions. 
However, for central Pb+Pb
collisions, the model predicts more narrow y - spectrum and less
steep $M_T$ - distribution: the model overestimates by 30-50\% the
$\phi$ yield at low $M_T-m_{\phi} < 0.15$ GeV$/c^2$ and
underestimates by 20-40\% that at higher $0.6 < M_T-m_{\phi} <
1.1$ GeV$/c^2$. The predicted value of the inverse slope parameter
$T_{\phi} = 232\pm2$ MeV of the $M_T$ - distribution turns out to
be much smaller than the experimental one, $T_{\phi} = 305\pm15$
MeV.\\ On the other hand, the model fits well the shape of the
$M_T$ - distribution for the dimuon channel of the $\phi$ and $\rho
+ \omega$ production in non-peripheral Pb+Pb collisions
\cite{ref7}. In the high $M_T$ - region (1.5 $< M_T <$ 3.2
GeV$/c^2$) the predicted values of $T_{\phi} = 220\pm7$ MeV and
$T_{\rho + \omega} = 220\pm3$ MeV agree with experimental ones.
The $({\rho + \omega})$ yield is also reproduced by the model
(when the FSI is included into the model),
unlike that of $\phi$ meson for which the predictions are 1.5-2
times lower than the data. \\ The situation worsens dramatically
for central Pb+Pb collisions (for the high $M_T$ region). The
model badly underestimates not only the $\phi$ yield (by a factor
$\sim$3), but also that for $({\rho + \omega})$ (by a factor
$\sim$2). This discrepancy is less
significant when comparing with the data \cite{ref6} which
concern the $\phi \rightarrow {K^+}{K^-}$ channel in the same
range of $M_T$.  \\ One can conclude, that the model fits,
irrespective to $N_{part}$, the $\phi$ total yield of $\phi$
(for pp, p+Pb, Pb+Pb), but fails at high $M_T$ region ($M_T >$ 1.5
GeV$/c^2$), especially for the  central Pb+Pb collisions for
which, however, the magnitude of the discrepancy with data is
rather different for different experiments \cite{ref6,ref7}.

\noindent In conclusion, the LUCIAE model reproduces some
features of the light vector meson production in proton-proton,
proton-nucleus and nucleus-nucleus collisions at the SPS energies,
but fails in describing the whole totality of the available data.
Although the mechanism of the formation of multistring states in
nuclear collisions, incorporated into the LUCIAE model,
provides a strangeness enhancement, it turns out to be far from being
sufficient to fit the data on the high-$M_T$ $\phi$ meson yield in the
central heavy ion collisions. The final state interactions, introduced
into the model, result in a significant increase of the $\omega$ yield
and lead to a better agreement with the experimental data.

\newpage
\noindent
 {\Large{\bf Acknowledgement}}\\
\vskip 0.1cm \noindent The authors are grateful to A.Grigoryan for
useful discussions and comments. The work is partially supported 
by Calouste Gulbenkian Foundation and Swiss Fonds "Kidagan".


\newpage

\begin{figure}
\begin{center}
\resizebox{.8\textwidth}{!}{\includegraphics*[bb=5 30 600
380]{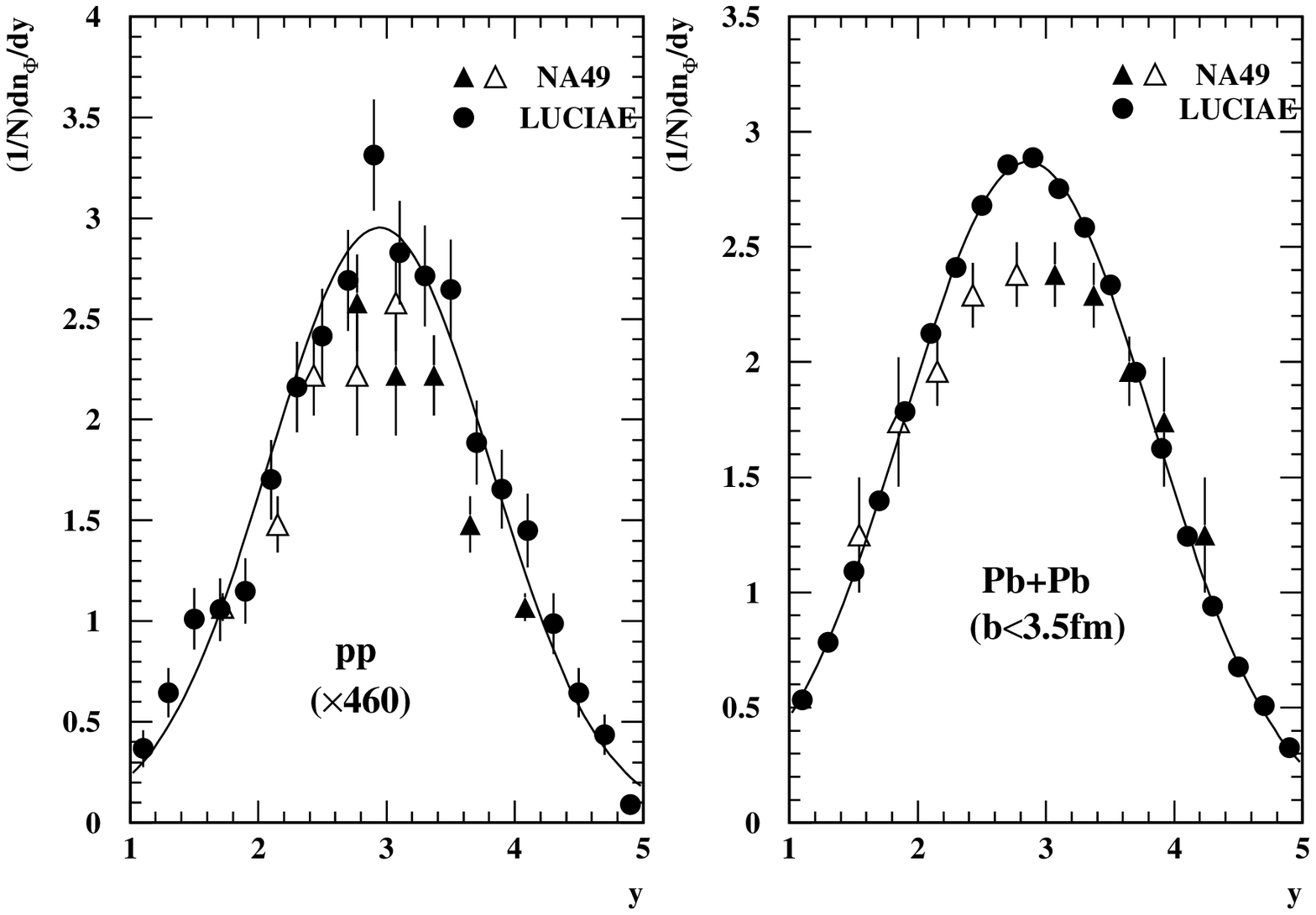}}
\caption{{ Rapidity distributions of $\phi$ mesons in pp and Pb+Pb
collisions. The curves are the result of the Gaussian fit of
the simulated data. }}
\end{center}

\begin{center}
\resizebox{.7\textwidth}{!}{\includegraphics*[bb=3 15 555
550]{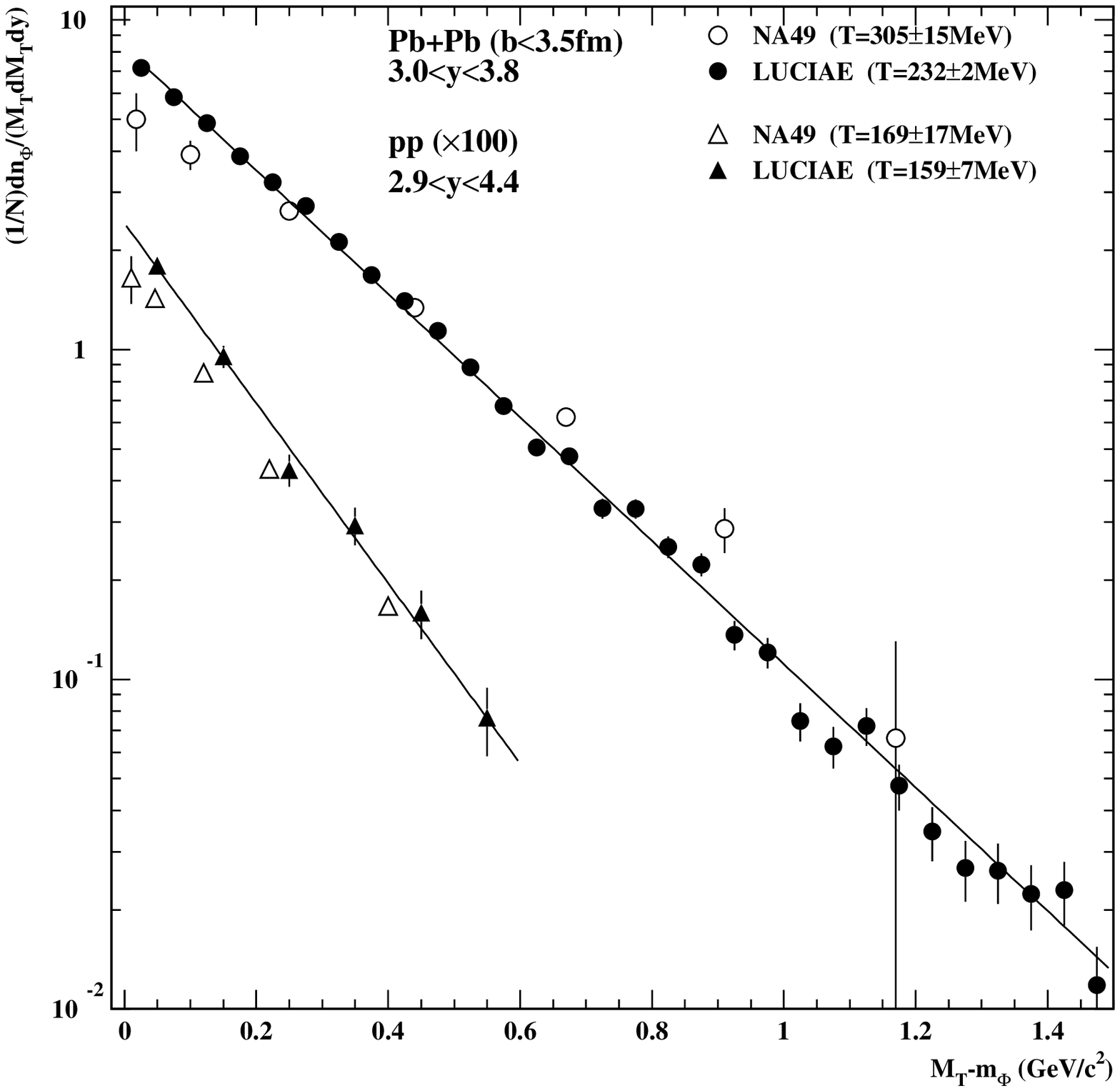}}
\caption{{ Transverse mass distribution of $\phi$ mesons in pp
and central Pb+Pb collisions. The curves are the result of the
exponential fit of the simulated data.}}
\end{center}
\end{figure}

\newpage

\begin{figure}
\begin{center}
\resizebox{.7\textwidth}{!}{\includegraphics*[bb=5 35 600
520]{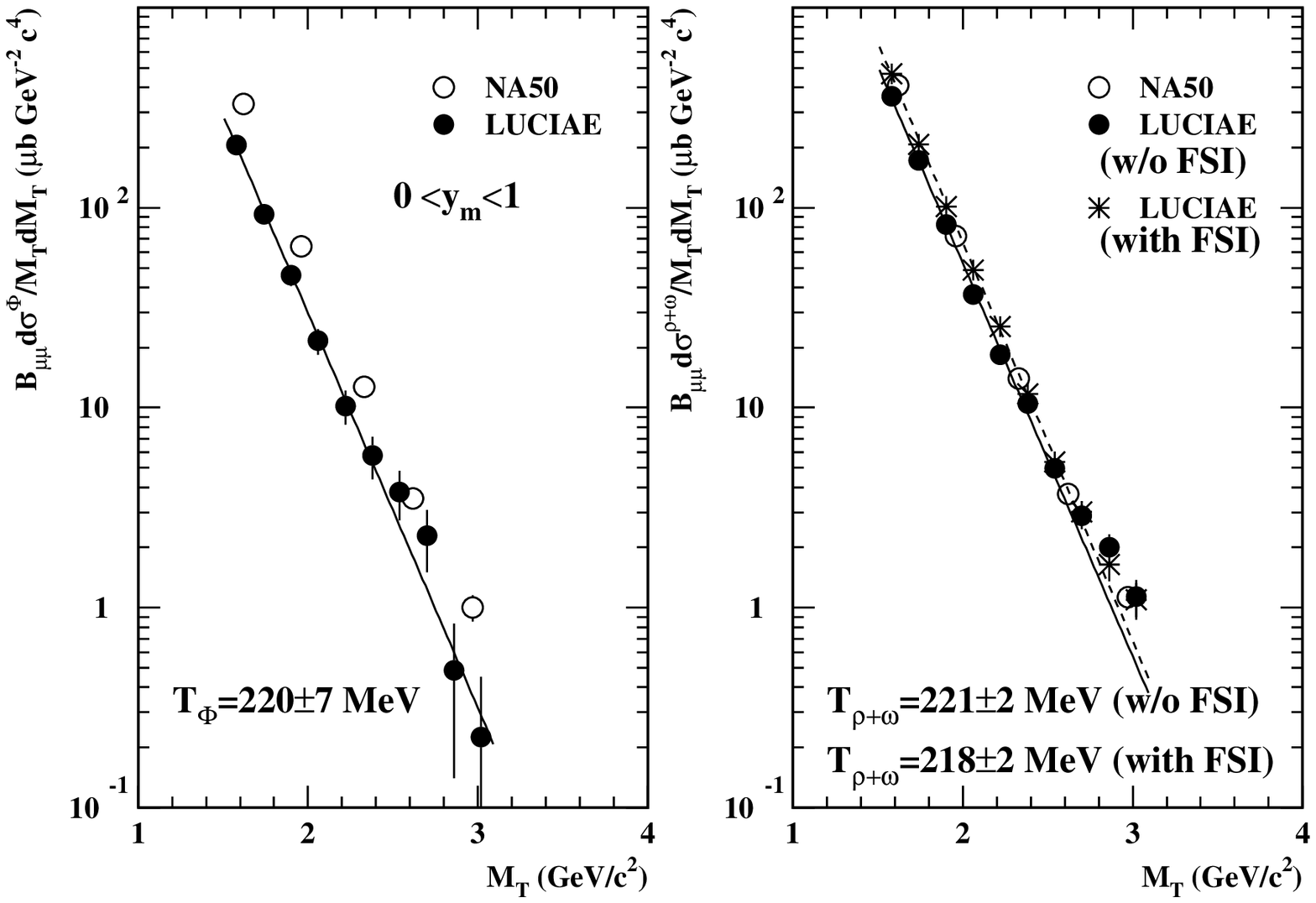}}
\caption{{Transverse mass distributions of $\phi$ and $\rho +
\omega$ in Pb+Pb collisions. The curves are the result of the
fit(see text) of the simulated data without (solid lines) and
with(dashed line) the FSI. }}
\end{center}

\vspace{0.5cm}

\begin{center}
\resizebox{.7\textwidth}{!}{\includegraphics*[bb=3 15 555
520]{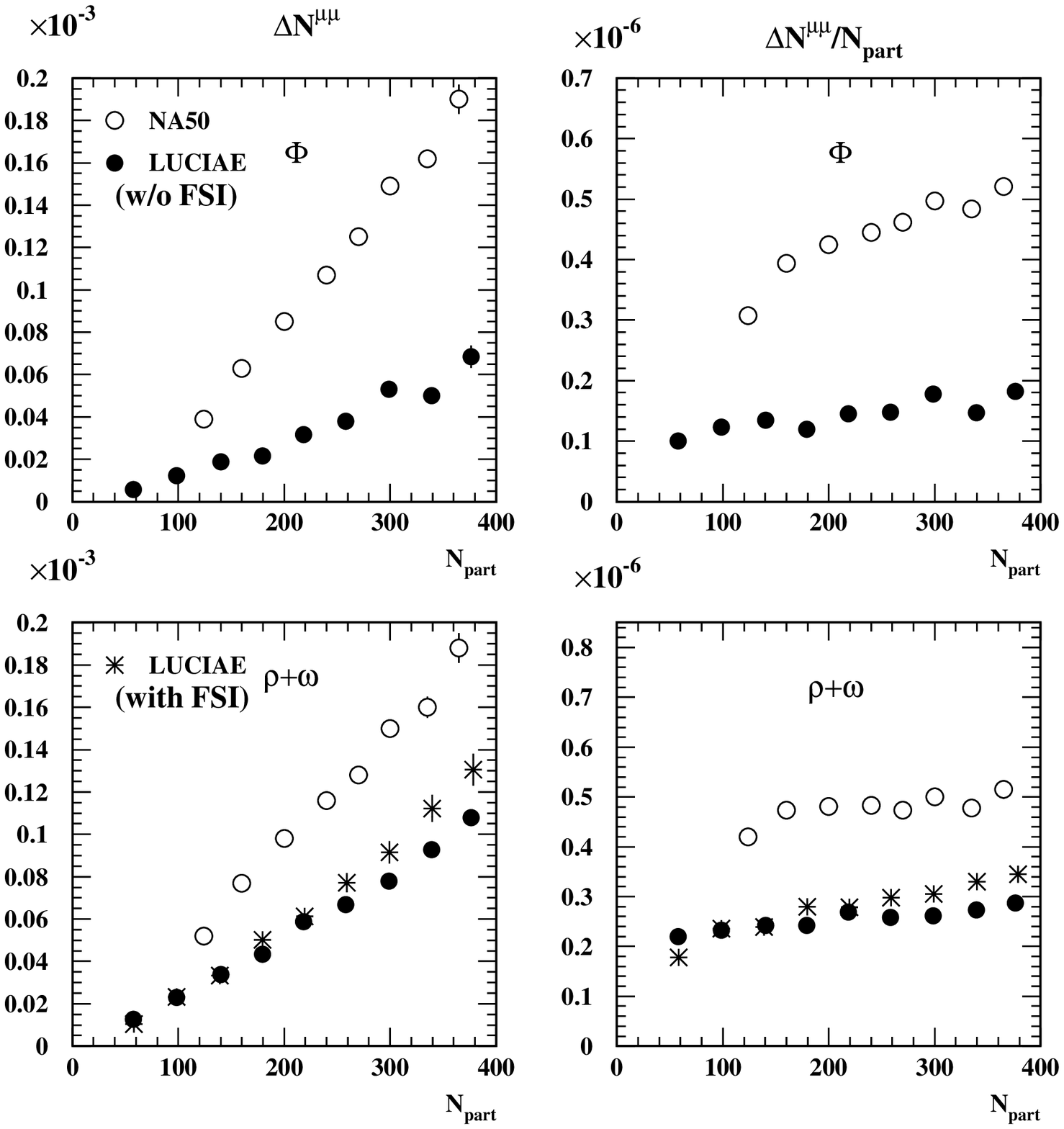}}
\caption{{Multiplicities $<n_{\phi}>_{\mu\mu}$ and $<n_{\rho +
\omega}>_{\mu\mu}$ (left panels) and multiplicities per
participant nucleon (right panels) in Pb+Pb collisions.}}
\end{center}
\end{figure}

\newpage

\begin{figure}
\begin{center}
\resizebox{.7\textwidth}{!}{\includegraphics*[bb=5 35 600
520]{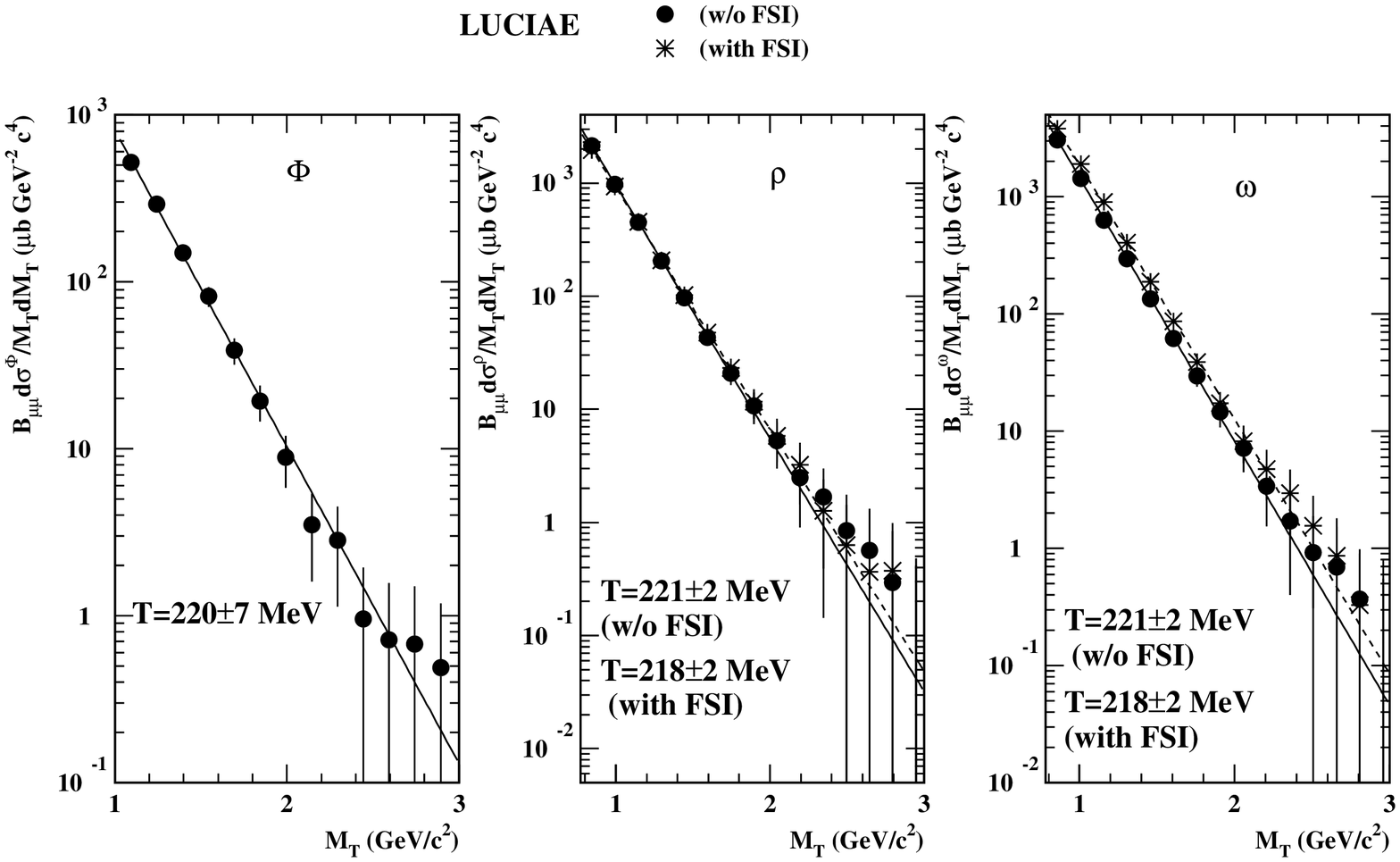}}
\caption{{Transverse mass distributions of $\phi$, $\rho$ and
$\omega$ in In+In collisions. The curves are the result of the
fit(see text) of the simulated data without (solid lines) and
with(dashed line) the FSI. }}
\end{center}
\end{figure}
\vspace{0.5cm}

\begin{figure}
\begin{center}
\resizebox{.7\textwidth}{!}{\includegraphics*[bb=3 15 555
520]{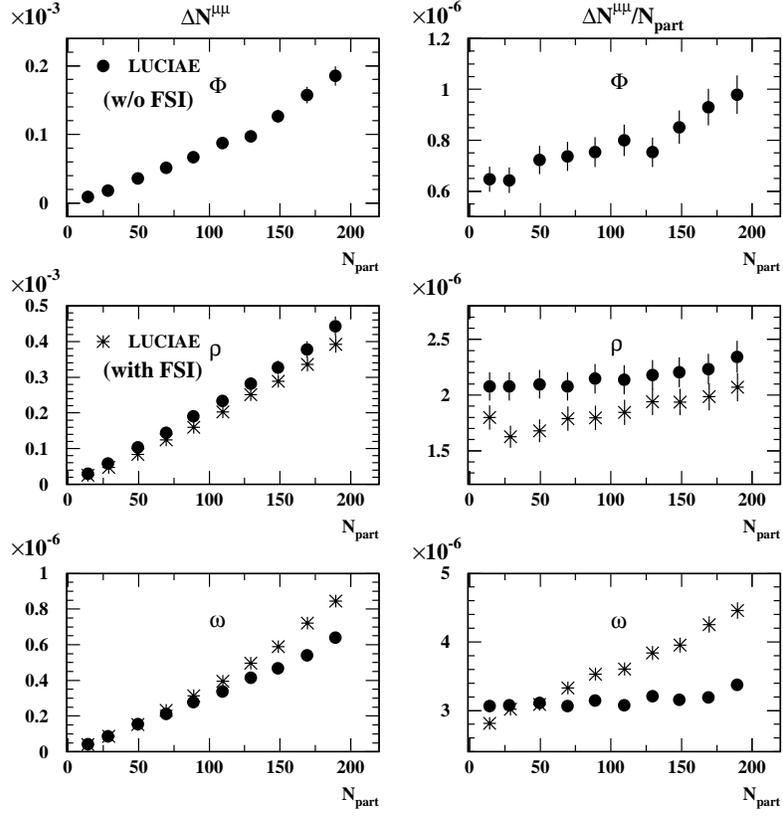}}
\caption{{Mean multiplicities $<n_{\phi}>_{\mu\mu}$,
$<n_{\rho}>_{\mu\mu}$ and $<n_{\omega}>_{\mu\mu}$
integrated over the whole $M_T$ - region (left panels) and those per
participant nucleon (right panels) in In+In collisions.}}
\end{center}
\end{figure}
\vspace{0.5cm}

\begin{figure}
\begin{center}
\resizebox{.7\textwidth}{!}{\includegraphics*[bb=3 15 555
520]{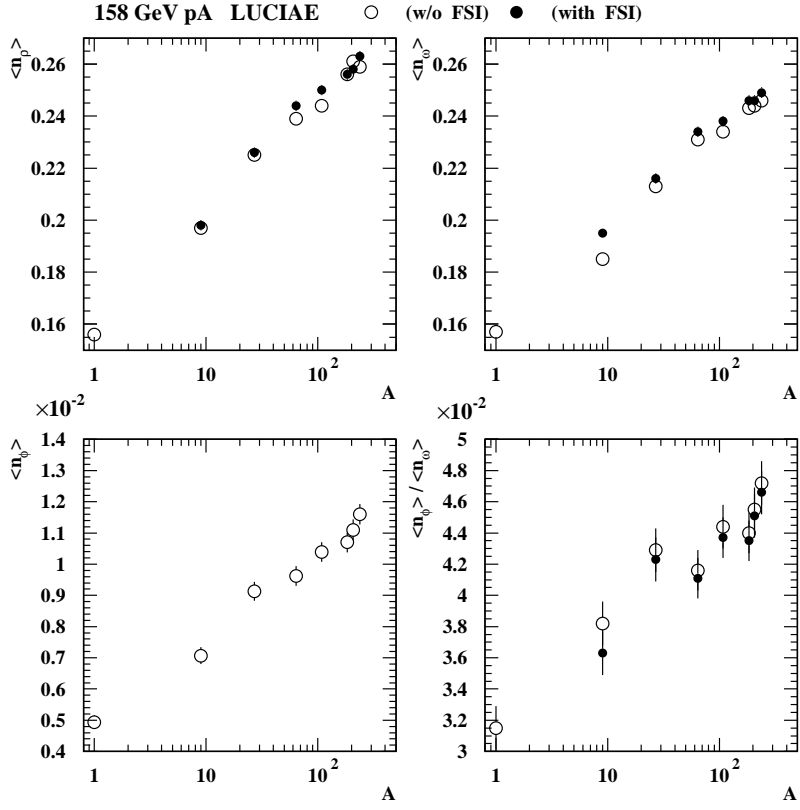}}
\caption{{The A-dependence of the mean multiplicities $<n_{\rho}>$,
$<n_{\omega}>$ and $<n_{\phi}>$ integrated over the whole $M_T$ - region
and the ratio $<n_{\omega}>/<n_{\phi}>$ in p+A collisions.}}
\end{center}
\end{figure}
\vspace{0.5cm}

\begin{figure}
\begin{center}
\resizebox{.7\textwidth}{!}{\includegraphics*[bb=3 15 555
520]{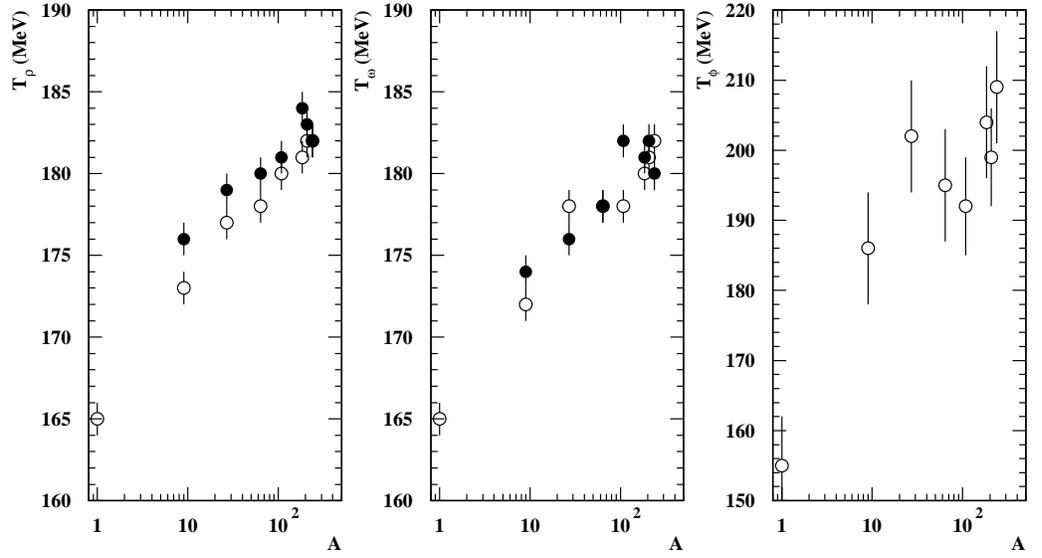}}
\caption{{The A-dependence of the inverse slope parameters $<T_{\rho}>$,
$<T_{\omega}>$ and $<T_{\phi}>$ of the $M_T$ - distributions
in p+A collisions.}}
\end{center}
\end{figure}

\end{document}